# Pushing the feature size down to 11 nm by interference photolithography with hyperbolic metamaterials under conventional UV light source

Xuefeng Yang [1,*], Shuxia Zhang[1], Baoji Wang[1], Xiaolin Cai[1], Xiaohua Li[1], Weiyang Yu[1], Qin Wang[1], & Zhongliang Lu[2]

[1]*School of Physics and Electronic Information Engineering, Henan Polytechnic University, Jiaozuo, 454003, China*
[2]*School of Safety Science and Engineering, Henan Polytechnic University, Jiaozuo, 454003, China*
*\* xfyang@hpu.edu.cn*

**Abstract:** Limited by the cost and complexity, superresolution lithography is hard to achieve through the traditional interference lithography. We here developed the plasmonic interference lithography technique by using a hyperbolic metamaterials (HMMs) / photoresist / metal plasmonic waveguide to push the feature sizes theoretically down to 16 nm and even to 11 nm at the wavelength of 365 nm with TM polarization. The waveguide based on the proposed HMMs can support high-$k$ mode for superresolution lithography. Furthermore, plasmonic mode supported in the proposed lithography structure can be tailored by dimension of HMM and permittivity of the materials, which makes it possible to get higher resolution pattern under conventional UV light. Our findings open up an avenue to pushing the nanolithography node towards 10 nm for low-cost and large area fabrication under conventional UV light source.

**Introduction**

The rapid progress in the nanoscale science and technology has increased the demand for fabrication of nanoscale optical and electronic devices. The nanolithography technologies have received a widely attention because of the demand for high resolution and cost effective. However, the traditional interference lithography techniques suffer the resolution barrier due to the diffraction limit of light. To improve the interference lithography resolution, one straightforward way is to using shorter wavelength light such as the deep ultraviolet light (DUV) [1] or extreme ultraviolet light (EUV) [2], and the other way is immersing the system in a higher refractive index material or fluid to increasing the numerical aperture of imaging system. But these will increase instrument complexity and the corresponding cost. Surface plasmons (SPs) [3] is a fascinating candidate light source for its wavelength is much smaller than that of light in free space at the same frequency. And because of such, the plasmonic nanolithography was demonstrated to improve the photolithography resolution in fabricating periodic pattern with interference lithography [4-7] or arbitrary nanopattern with imaging lithography [8-11]. In order to further enhance the resolution, hyperbolic metamaterials (HMMs) consisting of dielectric-metal multilayer structures were subsequently introduced in the nanolithography system [12-20]. This is for the HMMs can be specifically tailored to satisfy high-$k$ modes transmitted through to photoresist layer, and the dispersion of HMMs can be engineered by the permittivity and thickness of HMMs. The half-pitch resolution of 22 nm would be obtained with hyperbolic metamaterials [14], and the resolution was pushed down to 15 nm through employing the asymmetric slot SPP mode in a metal-insulator-metal (MIM) plasmonic waveguide under 193 nm illumination [21]. However, the 193 nm exposure system is too expensive and complex to be used for the plasmonic lithography.

It is good news that researchers had experimentally realize 22 nm half-pitch with imaging lithography [22] and plasmonic interference lithography [23] with conventional UV light. The next effort will be focused on breaking plasmonic lithography node down to 16 nm even to 11 nm under conventional UV light source. Higher resolution pattern need higher-$k$ mode excited in the lithography structure, the HMMs is a very good choice for tailoring the transverse wavenumber in the lithography structure. Ref.16 and Ref.19 demonstrate fabricating large area pattern with half-pitch 45 nm and 35 nm with the help of HMMs. However, excessive restraint the zero-order diffraction of mask need more layers of metal/dielectric multilayer and because of the loss characteristics of material, higher-$k$ mode wave would also be restrained. This will limit the resolution of lithography. In this work, through balancing the conflict between materials loss and high-$k$ mode wave, we developed the plasmonic interference lithography technique based on hyperbolic metamaterials (HMMs) / Photoresist (PR) / Al lithography structure to achieve the feature sizes theoretically down to 16 nm and even to 11 nm at the wavelength of 365 nm. This will make the pattern fabricating process economic and convenient.

**Structure and Methods**

Figure 1 shows the schematic of our proposed plasmonic interference lithography structure assisted by HMMs. Above the Si wafer, photoresist film and reflective Al film constitute the lithography structure. The Al grating, Al spacer and SiO$_2$/Al HMMs constitute the mask, which is fabricated on the quartz with the help of adhesive. The optical exposure system can be designed in a specially stage to ensure the conformal contact between the photomask and the lithography structure. The proposed HMMs in our work support high-$k$ spatial frequency transmitted to photoresist layer.

The high-$k$ mode of evanescent waves would be excited in the proposed HMMs/PR/Al waveguide. When the high-$k$ mode matched with the first order diffraction wave of the subwavelength grating, they would resonant with each other and enhance the electric field of interference pattern in the photoresist zone.

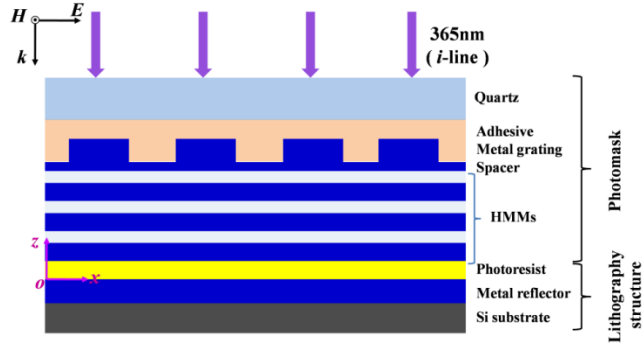

Figure 1. Schematic of the plasmonic interference lithography with hyperbolic metamaterials.

The dielectric-metal HMMs we proposed in this work (see the Fig.1) is characterized by its relative permittivities $\varepsilon_\perp < 0$ and $\varepsilon_{//} > 0$, that are perpendicular and parallel to the anisotropy axis $z$, respectively. The relative permittivity tensor in the case of uniaxial anistropy:

$$\varepsilon = \begin{bmatrix} \varepsilon_{xx} & 0 & 0 \\ 0 & \varepsilon_{yy} & 0 \\ 0 & 0 & \varepsilon_{zz} \end{bmatrix} \quad (1)$$

where the in-plane component $\varepsilon_{xx} = \varepsilon_{yy} = \varepsilon_\perp$, the out-of-plane component $\varepsilon_{zz} = \varepsilon_{//}$, and $\varepsilon_{//} \cdot \varepsilon_\perp < 0$. In this letter, we consider only electric uniaxial media and suppose its relative permeability $\mu_\perp = \mu_{//}$ simply reduces to the unit tensor. For transverse magnetic (TM) polarized wave, the dispersion relation of hyperbolic media can be described by

$$\frac{k_x^2 + k_y^2}{\varepsilon_{//}} + \frac{k_z^2}{\varepsilon_\perp} - k_0 = 0 \tag{2}$$

where $k_x, k_y$, and $k_z$ are the $x$, $y$, and $z$ components of the wavevector, respectively. $k_0 = \omega/c$ is the free-space wavenumber, $\omega$ is the wave frequency and $c$ is the speed of light in free space. According to the effect medium theory, the effective permittivity components for propagating perpendicular and parallel to the axis of anisotropy can be described as

$$\varepsilon_{xx} = \varepsilon_{yy} = \varepsilon_\perp = \frac{\varepsilon_m d_m + \varepsilon_d d_d}{d_m + d_d}$$

$$\varepsilon_{zz} = \varepsilon_{//} = \frac{d_m + d_d}{d_m/\varepsilon_m + d_d/\varepsilon_d} \tag{3}$$

where $\varepsilon_m$ and $\varepsilon_d$ are the permittivity of metal and dielectric, $d_m$ and $d_d$ are the thickness of metal and dielectric layer.

**Results and Discussion**

For $SiO_2$/Al HMMs proposed in our work, Figure 2 presents the relationship of effective permittivity with the metal thickness ratio ($\rho$) at the wavelength of 365 nm. It appears the type II HMM property when $\rho$ within the range from 0.2 to 0.8, which would be applied in the deep subwavelength plasmonic interference lithography.

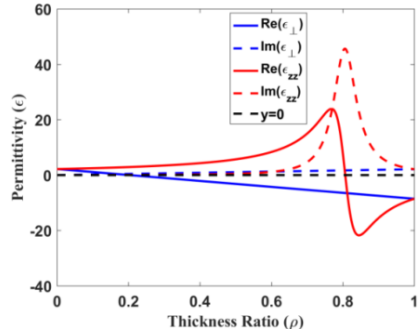

Figure 2. The relationship between effective permittivity and metal thickness ratio $\rho$ of $SiO_2$/Al HMMs at the wavelength of 365 nm.

In order to get the optimal thickness of the proposed HMMs to meet super-resolution nanolithography, the optical transmission function of the proposed HMMs versus the period of HMM and the transverse wavenumber has calculated through rigorous coupled wave analysis method (RCWA) with the metal film thickness ratio 0.6 at a wavelength of 365 nm, shown in Figure 3(a). Note that the thickness of photoresist layer and Al reflector layer are 16 nm and 30 nm, respectively. The permittivity of Al with 15 nm and 30 nm thick at 365 nm are corresponding to the experimental measured data of -8.5568+2.1102i and -9.1906+2.8292i, respectively [23]. The permittivity of photoresist and $SiO_2$ are 2.8643+0.1513i and 2.1742. Because of the aperture of the grating is small, we consider the grating and the Al spacer as a whole that the permittivity of them are both -9.1906 + 2.8292i. It can clearly see from Figure 3(a) that the high-$k$ mode would appear when the period of the proposed HMMs less than 30 nm. For 20 nm half pitch lithography, the transverse wavevector of waveguide mode should be about 365 $k_0$ / 80 = 4.56 $k_0$. For the convenience of experimentation, we chose the period of HMMs 25 nm. The optical transmission function of the proposed HMMs versus the metal film ratio of HMM and the transverse wavenumber is calculated and shown in Figure 3(b). High-$k$ mode (about 4.6 $k_0$) would appear within a broad metal fill ratio range from 0.15 to 0.75. This will make it more convenient for experimental realization.

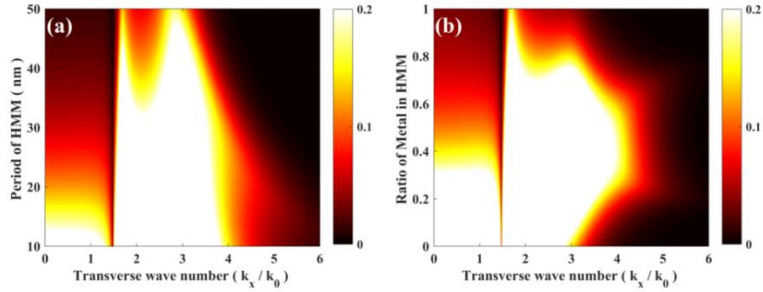

Figure 3. (a) The optical transmission function (OTF) of the proposed HMMs versus the period of HMM and the transverse wavenumber with metal film thickness ratio 0.6 at the wavelength of 365 nm. (b) The OTF of the proposed HMMs versus the metal film ratio of HMM and the transverse wavenumber with the period of HMMs 25 nm.

The finite-difference time-domain method is then performed for numerical analysis. The Al grating is 80 nm period with 20 nm thick and 18 nm aperture width and the Al spacer is 10 nm thick. Two layers of $SiO_2$ (10 nm) / Al (15 nm) HMMs is coating on the Al spacer. The thickness of PR and Al reflector layer are 16 nm and 30 nm, respectively. The simulated normalized electric-field intensity distribution in the $xz$ plane is given in Figure 4(a). Clearly the feature size of the interference pattern is down to 20 nm. The normalized total electrical field intensity distribution at $z$ = 4 nm (black, 1/4 *PR), 8 nm (blue, 1/2*PR), and 12 nm (red, 3/4*PR) above the photoresist/Al reflector interface is also shown in Figure 4(b). The intensity visibility $V = (I_{max} - I_{min}) / (I_{max} + I_{min}) \approx 0.42$, which satisfy the minimum contrast for positive photoresist in modern lithography fabrication processes.

From Figure 3(a), higher $k$ mode can be excited when the period of proposed HMMs decreases. In order to further increasing the resolution (i.e. 16 nm half pitch with transverse wavevector about 365 $k_0$ / 64 = 5.7 $k_0$) of interference lithography and without losing the experimental implementation, the period of HMMs is reduced to 20 nm. For the evanescent waves decay exponentially with the distance away from the propagating path, the thickness of photoresist is reduced to 12 nm for higher $k$ mode.

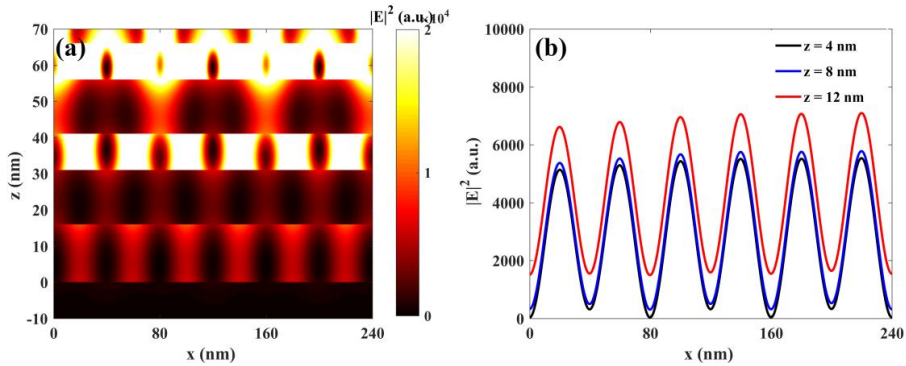

Figure 4. (a) Normalized electric-field intensity distribution with two layers of SiO2 (10 nm) / Al (15 nm) HMMs and 16 nm PR. (b) Normalized total electrical field intensity distribution at $z$ = 4 nm (black, 1/4 *PR), 8 nm (blue, 1/2*PR), and 12 nm (red, 3/4*PR) above the photoresist/Al reflector interface.

Figure 5(a) presents the optical transmission function of the 3 layers proposed HMMs versus the period of HMM and the transverse wavenumber with the metal film thickness ratio 0.6 at a wavelength of 365 nm. Note that the thickness of photoresist layer and Al reflector layer are 12 nm and 30 nm, respectively. It can clearly see that the higher $k$ mode would appear when the period of the proposed HMMs less than 20 nm. Then we calculate the optical transmission function of the proposed HMMs versus the metal film ratio of HMM and the transverse wavenumber with the period of HMMs 20 nm at a wavelength of 365 nm, which is shown in Figure 5(b). High-$k$ mode (about 5.7 $k_0$) would appear within a broad metal fill ratio range from 0.2 to 0.78. This will make it easier to implement in experimental realization.

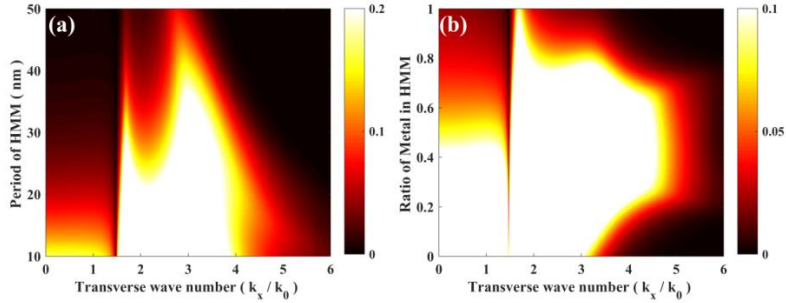

Figure 5. (a) The OTF versus the period of HMM and the transverse wavenumber for three layers of $SiO_2$ / Al HMMs with metal film thickness ratio 0.6 at the wavelength of 365 nm. (b) The OTF of the proposed HMMs versus the metal film ratio of HMM and the transverse wavenumber with the period of HMMs 20 nm.

The finite-difference time-domain method is then performed for numerical analysis at the wavelength of 365 nm with $p$-polarized. The Al grating is 64 nm period with 20 nm thick and 18 nm aperture width and the Al spacer is 10 nm thick. 3 layers of $SiO_2$ (5 nm) / Al (15 nm) HMMs is coating on the Al spacer. The thickness of PR and Al reflector layer are 12 nm and 30 nm, respectively. The simulated normalized electric-field intensity distribution in the $xz$ plane is given in Figure 6(a). Clearly the feature size of the interference pattern is down to 16 nm. The normalized total electrical field intensity distribution at $z = 3$ nm (black, 1/4 *PR), 6 nm (blue, 1/2*PR), and 9 nm (red, 3/4*PR) above the photoresist/Al reflector interface is also shown in Figure 6(b). The intensity visibility is about 0.43 which satisfy the minimum contrast for positive photoresist in modern lithography fabrication processes. It also demonstrates that even with a high zero-order diffraction wave exits in the waveguide, the first order interference lithography pattern with high uniformity will be also generated.

For further enhancement of the lithography resolution with the proposed structure, larger transverse wavevector need to be stimulated in the PR layer. From above discussion, the period of HMMs should be further decreased. However, larger wavevector propogates in PR layer would limit its propagating length and the PR layer should be decreased. On the other hand, higher index of the photoresist layer would also generate larger wavevector. For the above reasons, we chose 3 layers of GaN/Al as the proposed HMMs and hydrogen silsesquioxane (HSQ) as the photoresist. Note that, referring to the previous work [17,19], the experimental result agree well with simulational calculation even the permittivity of the Al is described with Drude model. The permittivity of GaN, Al, and HSQ are 7.043 [24], -19.66+4.4i, and 3.2 at 365 nm, respectively. The optical transmission function of the proposed HMMs versus the metal film ratio in HMM and the transverse wavenumber with the period of HMMs 10 nm is shown in Figure 7(a). The zoom of range from 8 $k_0$ to 9 $k_0$ is shown in Figure

7(b). It clearly shows that the transverse wavevector can be excited up to 8.3 $k_0$ (about 44 nm grating) within the range of metal thickness ratio from 0.3 to 0.6.

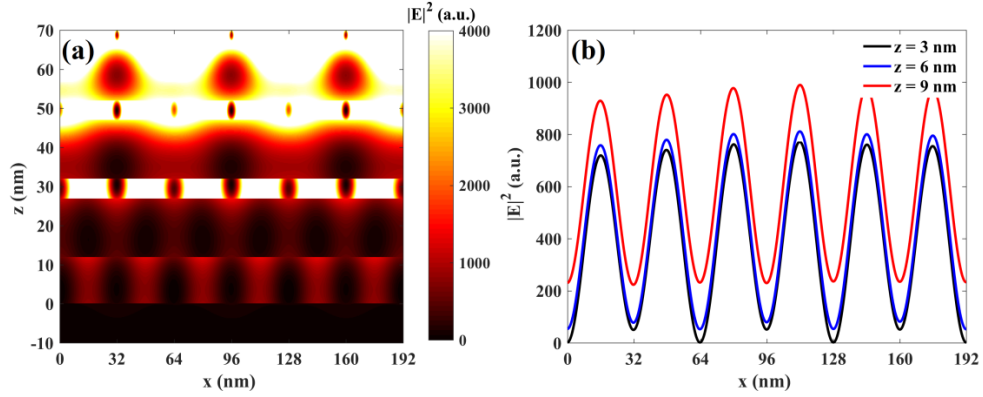

Figure 6. (a) Normalized electric-field intensity distribution with three layers of $SiO_2$ (5 nm) / Al (15 nm) HMMs and 12 nm PR. (b) Normalized total electrical field intensity distribution at z = 3 nm (black, 1/4 *PR), 6 nm (blue, 1/2*PR), and 9 nm (red, 3/4*PR) above the photoresist/Al reflector interface.

The simulated normalized electric-field intensity distribution in the xz plane is given in Figure 7(c). The Al grating is 44 nm period with 24 nm thick and 12 nm aperture width. The Al spacer is 10 nm thick. 3 layers of GaN (4.4 nm) / Al (5.6 nm) HMMs is coating on the Al spacer. The thickness of PR and Al reflector layer are 8 nm and 30 nm, respectively. Clearly the feature size of the interference pattern is down to 11 nm with intensity visibility about 0.31, which satisfy the minimum contrast for negative photoresist HSQ in fabrication processes. The normalized total electrical field intensity distribution at z = 2 nm (black, 1/4 *PR), 4 nm (blue, 1/2*PR), and 6 nm (red, 3/4*PR) above the photoresist/Al reflector interface is also shown in Figure 7(d).

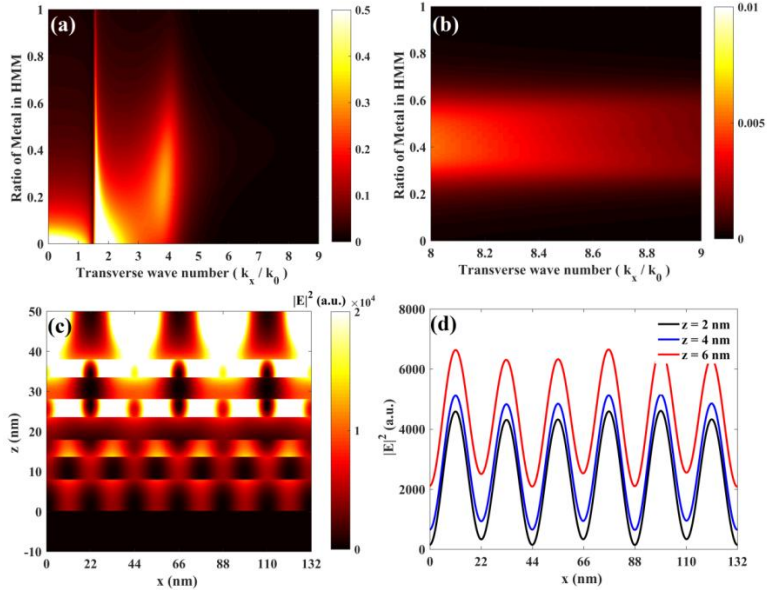

Fig 7. (a) The OTF versus the metal film ratio of HMM and the transverse wavenumber for three layers of GaN/Al HMMs with the period of HMMs 10 nm at the wavelength of 365 nm. (b) The zoom of Figure 7(a) from $8k_0$ to $9k_0$. (c) Normalized electric-field intensity distribution

with three layers of GaN (4.4 nm) / Al (5.6 nm) HMMs and 8 nm PR. (d) Normalized total electrical field intensity distribution at z = 2 nm (black, 1/4 *PR), 4 nm (blue, 1/2*PR), and 6 nm (red, 3/4*PR) above the photoresist/Al reflector interface.

Though the exciting high lithography resolution is demonstrated with our proposed structure at the conventional UV light source 365 nm, actual implementation will be a challenge for structure fabrication, especially for the 11 nm node case. The quality features of films such as smoothness, hardness, and stickiness should be fully considered in experiment. However, Ref. 17 demonstrates 6 nm Al film can be well control to coat in the $Al/Al_2O_3$ multilayer, it will guide us for the coating of GaN (4.4 nm) / Al (5.6 nm) HMMs in our proposed case.

## Conclusion

In this work, we developed the plasmonic interference lithography technique by using a hyperbolic metamaterials (HMMs) / photoresist / metal plasmonic waveguide. Through balancing the conflict between materials loss and high-$k$ mode wave in the proposed HMM waveguide, we pushed the feature sizes theoretically down to 16 nm and even to 11 nm at the wavelength of 365 nm. We believe that our findings would open up an avenue to pushing the nanolithography node towards 10 nm for low-cost and large area fabrication under conventional UV light source.

## Acknowledgments
This work was supported by National Natural Science Foundation of China (NSFC) (U1604133, 12074102 and 11804082), Foundation of Henan Educational Committee (No. 20A140013).

## Disclosures

The authors declare no conflicts of interest.